\documentclass[aps,showpacs,pra,twocolumn,superscriptaddress,a4paper]{revtex4-1}

\usepackage{amsmath}
\usepackage{amsthm}
\usepackage{amssymb}
\usepackage{amscd,bm}
\usepackage{wasysym}
\usepackage[ansinew]{inputenc}
\usepackage[T1]{fontenc}
\usepackage{ae,aecompl}
\usepackage{hyperref}
\usepackage{array}
\newcolumntype{L}{>{$}l<{$}} 

\usepackage{float}

\usepackage [english]{babel}
\usepackage{cleveref}
\crefname{figure}{Fig.}{Figures} 
\Crefname{figure}{Figure}{Figures} 
\crefname{section}{Sec.}{Sections} 
\Crefname{section}{Section}{Sections} 
\crefname{table}{Tab.}{tables} 
\crefname{equation}{Eq.}{Equations} 
\Crefname{equation}{Equation}{Equations} 

\usepackage{graphicx}

\newcommand{\be}{\begin{equation}}
\newcommand{\ee}{\end{equation}}

\begin{document}

\title{Modeling long correlation times using additive binary Markov chains: applications to wind generation time series}
\author{Juliane Weber}
\email{ju.weber@fz-juelich.de}
\affiliation{Forschungszentrum J\"ulich, Institute of Energy and Climate Research -- Systems Analysis and Technology Evaluation (IEK-STE), 52425 J\"ulich, Germany}
\affiliation{University of Cologne, Institute for Theoretical Physics, Z\"ulpicher Str.~77, 50937 Cologne, Germany}

\author{Christopher Zachow}
\affiliation{University of Cologne, Institute for Theoretical Physics, Z\"ulpicher Str.~77, 50937 Cologne, Germany}

\author{Dirk Witthaut}
\affiliation{Forschungszentrum J\"ulich, Institute of Energy and Climate Research -- Systems Analysis and Technology Evaluation (IEK-STE), 52425 J\"ulich, Germany}
\affiliation{University of Cologne, Institute for Theoretical Physics, Z\"ulpicher Str.~77, 50937 Cologne, Germany}

\date{\today }

\begin{abstract}
Wind power generation exhibits a strong temporal variability, which is crucial for system integration in highly renewable power systems. Different methods exist to simulate wind power generation but they often cannot represent the crucial temporal fluctuations properly. We apply the concept of additive binary Markov chains to model a wind generation time series consisting of two states: periods of high and low wind generation. The only input parameter for this model is the empirical autocorrelation function. The two state model is readily extended to stochastically reproduce the actual generation per period. To evaluate the additive binary Markov chain method, we introduce a coarse model of the electric power system to derive backup and storage needs. We find that the temporal correlations of wind power generation, the backup need as a function of the storage capacity and the resting time distribution of high and low wind events for different shares of wind generation can be reconstructed. 
\end{abstract}

\maketitle

\section{Introduction}

Mitigation of climate change requires the decarbonization of the energy system \cite{Paris15,roge15,Rogelj16}. Power plants based on fossil fuels must be replaced by renewable sources such as wind and solar power. These technologies have shown remarkable progress in the last decades but the integration into the energy system represents a huge challenge due to their strong intermittency \cite{Sims11,Bloo16,olau16,Schiel17,Scha17,Schmie16,Scha18}. Generation and load have to be balanced at every instance of time. Thus, for a system dominated by fluctuating renewable sources, large amounts of backup and storage are required in order to always guarantee this power balance \cite{Heid10,Rasm12,diaz12,Elsn15}.

Wind power, as one of the main renewable energy sources, exhibits temporal correlations on different time scales from seconds \cite{Mila13,anva16} over weeks and years \cite{Schlachtb16} to decades \cite{Bloo16}. Periods with below (and above) average wind power generation can last up to weeks \cite{Cann15,Weber17}. Furthermore, the generation can vary significantly on hourly timescales, even when aggregated over large spatial scales \cite{Pesc15,grams17,2017Wohland,Bandi17}. These fluctuations have to be accounted for when designing a future energy system. 

High amounts of backup and storage infrastructures will be necessary in energy systems with a high share of wind power to provide energy in times of low renewable generation \cite{Heid10,Rasm12,Jens14,Staf17}. In order to quantify the need for backup and storage, either long-term measurements or reliable models of wind power generation are needed. As long-term measurements are rare, it is important to develop models which are able to represent the temporal fluctuations of wind power generation properly.

Stochastic models are commonly used to simulate wind speed and wind generation data \cite{Brok09,Papa08,Hagh13,Pesc15,Cara13a,Cara13}. Typical models include Autoregressive Moving Average (ARMA) and Markov chain models. ARMA models require a high amount of model parameters. Simple first-order Markov models often cannot reproduce the temporal characteristics of wind power generation, in particular long correlation times, and therefore they are of very limited use for energy system analyses \cite{Brok09}. Higher order Markov models require a huge amount of input data and are thus impractical \cite{Pesc15,Brok09,Papa08}. Additionally, for ordinary Markov models the data has to be discretized and the model output depends on the exact discretization scheme \cite{Pesc15}. 

Stochastic models for wind power generation should reproduce the main temporal characteristics but remain simple enough for practical applications. We introduce a simple, two-state stochastic model based on additive binary Markov chains developed in \cite{Usat03,usat03a,Meln06,Meln06b}. Additive Markov chains are an efficient tool to simulate time series with long-range correlations because the transition probability can be expressed as a sum of functions each depending on one of the previous states (memory functions). Strong analytic results exist for the case of binary time series \cite{Usat03,usat03a,Meln06,Meln06b}. In this case, the memory functions can be derived from the empirical autocorrelation function of the time series straightforwardly. Additionally, the model depends on the memory length which determines the maximum time lag taken into account to derive the memory functions. In this paper, we thus consider a simplified, binary wind generation time series, distinguishing only between two different types of system states: Scarcity in times of low, and oversupply in times of high wind generation. Already for this simple model, in which the wind power generation can assume only two values for the two states, we obtain a fairly good stochastic model. Our purpose is to reconstruct the temporal aspects of a wind power time series as they are crucial for system operation, especially for high shares of renewables combined with large storage facilities. Therefore, we evaluate our model by analyzing the temporal correlations, the backup need as a function of the storage capacity and the resting time distributions of scarcity and overproduction events for different shares of wind generation. 

The paper is organized as follows: We introduce the additive binary Markov chain method (\cref{sec:Methods_Additive_Markov}), discuss the available wind power data and present a coarse grained model of the electric power system used to test the stochastic model (\cref{sec:Methods_Application_to_Wind}). Subsequently, we compare the characteristics of the input and the synthetic data in order to evaluate our model (\cref{sec:Results}). In a further step, we present a way to extend our model to non-binary wind generation data (\cref{sec:tauA}). Finally, we close with some concluding remarks (\cref{sec:Conclusion}).

\section{Additive binary Markov chains} \label{sec:Methods_Additive_Markov}

A $1$-step Markov chain of a homogeneous stochastic process with a discrete number of states is defined as a sequence of random variables $X_1$, $X_2$, $X_3, \dots $ in which the conditional probability for the future state $X_t=x_t$ is determined entirely by the knowledge of the previous state $X_{t-1}=x_{t-1}$ \cite{VanK92,gardiner85},
\be
\begin{aligned}[c]
P(X_t &=x_t | X_{t-1} =x_{t-1}, \dots , X_{0}=x_{0}) \\
&= P(X_t=x_t | X_{t-1}=x_{t-1}).
\end{aligned}
\label{eq:1-state_Markov}
\ee
\noindent For practical applications, this conditional probability is reconstructed from measured data by discretizing the data such that the state space of the random variable becomes finite.  

To include memory, \cref{eq:1-state_Markov} can be generalized to an $N$-step Markov chain for which the future state depends on $N$ previous states,
\be 
\begin{aligned}[c]
P(X_t &=x_t | X_{t-1} = x_{t-1}, \dots , X_{0}=x_{0}) \\
&= P(X_t=x_t | X_{t-1}=x_{t-1}, \dots , X_{t-N}=x_{t-N}).
\end{aligned}
\label{eq:N-state_Markov}
\ee

\noindent However, as \citet{Brok09} point out, ``the problem with higher order Markov models is that there are $K^N$ states, where $K$ is the number of discretized wind powers and $N$ is the order of the model, which is intractable for large $N$" and therefore for long memory times.

Assuming that the influence of previous states on the future state is additive, \cref{eq:N-state_Markov} can be expressed as a sum of functions $f$, each depending on one of the previous states
\be
\begin{aligned}[c]
P(X_t &=x_t | X_{t-1} =x_{t-1}, \dots , X_{t-N}=x_{t-N}) \\
&= \sum_{r=1}^{N} f(x_t, x_{t-r}, r).
\end{aligned}
\label{eq:add_Markov}
\ee

Strong analytical results are available for additive binary Markov chains, i.e.~stochastic processes which assume only two states $a_t = \{0,1\}$ \cite{Usat03,usat03a,Meln06}. In this case, the conditional probability can be simplified to \cite{Usat03,usat03a,Meln06} 
\be 
P(a_t=1 | T_{N,t}) =  \sum_{r=1}^{N} f(a_{t-r}, r),
\label{eq:add_markov_general}
\ee
\noindent with $T_{N,t} = a_{t-1}, a_{t-2}, \dots, a_{t-N}$. The function $f(a_{t-r}, r)$ describes the contribution of $a_{t-r}$ to the conditional probability of $a_t$ taking the value $a_t=1$ at time step $t$. The value $N$ can be interpreted as the memory length of the Markov chain.

\citet{Meln06} show that \cref{eq:add_markov_general} can be rewritten in the following, simpler form
\be 
P(a_t=1 | T_{N,t}) = \langle a \rangle + \sum_{r=1}^{N} F(r) \, (a_{t-r} - \langle a \rangle),
\label{eq:add_markov}
\ee
\noindent with $\langle a \rangle$ being the average over the whole sequence and $F(r)$ being a memory function, which describes the strength of the influence of the previous value $a_{t-r}$ on $a_t$.

The following relationship between the memory function $F(r')$ and the autocovariance function $K(r)$ exists \cite{Meln06}:
\be
K(r) = \sum_{r'=1}^{N} F(r') \, K(r-r'), \quad r \geq 1,
\label{eq:lin_eq_Fr}
\ee
\noindent with 
\par\nobreak\noindent 
\begin{align}
K(r) & = \langle a_t \, a_{t+r} \rangle -  \langle a \rangle^2, \label{eq:Kr} \\
K(0) & = \langle a \rangle \, (1-  \langle a \rangle), \\
K(r) & = K(-r).
\end{align}
\noindent \Cref{eq:lin_eq_Fr} is a system of linear equations and can be solved for $F(r')$ straightforwardly. Thus, it is sufficient to know the autocovariance function of a time series in order to reconstruct the memory function which fully characterizes the stochastic process. This memory function can then be inserted into \cref{eq:add_markov} to compute synthetic time series for Monte Carlo simulations.

\section{Application to wind generation data} \label{sec:Methods_Application_to_Wind}
We apply the additive binary Markov model introduced in \cref{sec:Methods_Additive_Markov} to wind generation time series. One prime benchmark for such a model is whether important characteristics of power system operation can be reproduced. For this purpose, we introduce a simple, coarse-grained model of the electric power system in \cref{sec:Methods_Power_System} to derive backup and storage needs which crucially depend on the temporal characteristics of wind power generation, especially for high shares of wind power \cite{Heid10,Rasm12}. Afterwards, we present the empirical data used as input for the construction of the model (\cref{sec:Methods_Wind}) and explain the mapping of the continuous wind generation time series to a binary time series (\cref{sec:binarization_of_data}). Finally, we explain the initialization of our model in \cref{sec:initialization}.

\subsection{Model of the electric power system} \label{sec:Methods_Power_System}

At every time step $t$, the generated and consumed electric power must be balanced \cite{Heid10,Rasm12,Rodriguez2014}:
\be
R(t) + B(t) = \Delta(t) + L(t) + C(t).
\label{eq:power_balance}
\ee
\noindent Here, $R(t)$ is the renewable power generation time series which we want to model using the additive binary Markov chain method, and $L(t)$ is the load. If the load exceeds the generation, i.e.~$L(t)>R(t)$, the missing energy must be provided by storage facilities ($\Delta(t) < 0$) or by conventional backup power plants ($B(t) > 0$). The backup need can be interpreted as the aggregated amount of power required from dispatchable power plants. If, on the other hand, the generation exceeds the load, i.e.~$R(t)>L(t)$, excess energy can either be stored ($\Delta(t) > 0$) or the generation from renewables has to be curtailed ($C(t) > 0$). Hence, the storage filling level $S(t)$ evolves according to
\be
S(t+T) = S(t) + \Delta(t) \, \cdot \, T,
\label{eq:storage}
\ee
\noindent where $T$ denotes the duration of one time step (here: 1 hour). Storage facilities are limited such that the storage filling level must satisfy $0 \leq S(t) \leq S_{\rm max}$. 

The energy that has to be provided by conventional backup power plants is minimized for a storage-first strategy \cite{Rasm12}. Thus, $\Delta(t)$ can be derived in the following way: 
\par\nobreak\noindent 
\begin{align}
\Delta(t) = \begin{cases}
			\min[R(t) - L(t); \frac{S_{\rm max} - S(t)}{T}] &{\rm if} \quad R(t) > L(t) \\
			-\min[L(t) - R(t); \frac{S(t)}{T}] &{\rm if} \quad R(t) < L(t).
			\end{cases}
\end{align}
\noindent Accordingly, $B(t)$ and $C(t)$ read:
\par\nobreak\noindent 
\begin{align}
B(t) = 
\begin{cases}
L(t) - R(t) + \Delta(t) & \quad {\rm if} \quad L(t) > R(t) \\
0 & \quad {\rm else} \\
\end{cases}\\
C(t) = 
\begin{cases}
R(t) - L(t) - \Delta(t) & \quad {\rm if} \quad R(t) > L(t) \\
0 & \quad {\rm else}. 
\end{cases}
\end{align}

A central question for energy system operation is how much storage and backup are needed in a highly renewable power system \cite{Heid10,Rasm12,Jens14}. We therefore use the temporal mean of the backup energy $\langle B \rangle$ as a function of the storage capacity $S_{\rm max}$ as a benchmark for the developed Markov chain model. For simplicity, we normalize the average backup energy by the average load $\langle L \rangle$ to obtain the average share of energy that has to be provided by backup power plants:
\be
E = \frac{\langle B \rangle}{\langle L \rangle}.
\ee

\begin{figure*}[tb]
\centering
\includegraphics[width=0.99\textwidth]{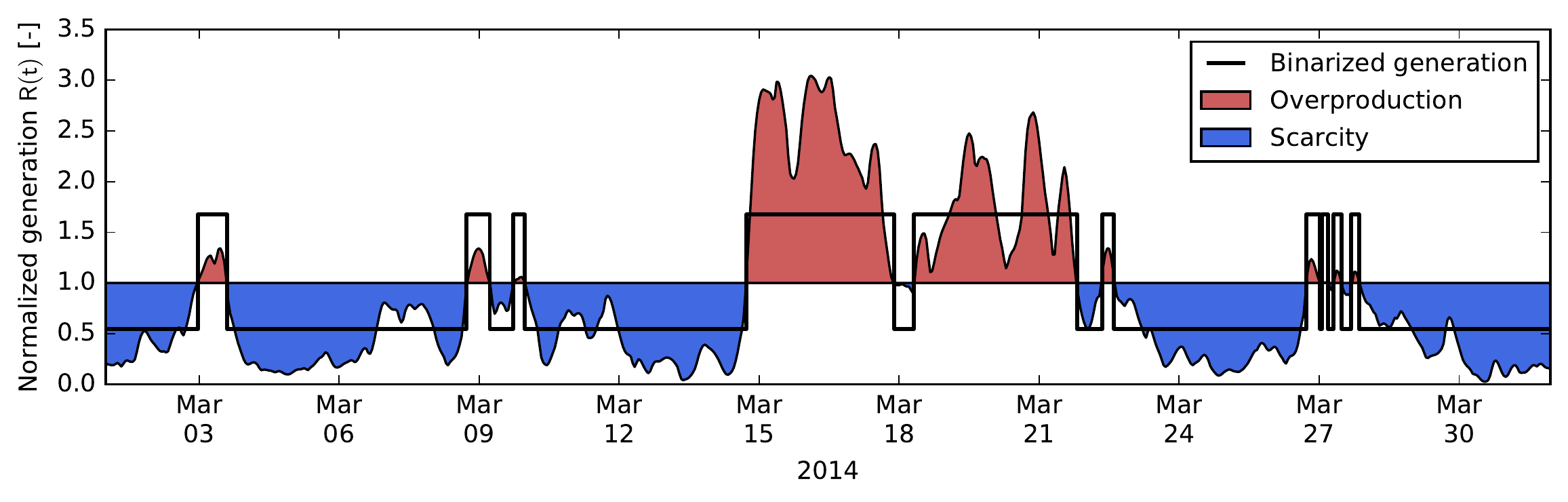}
\caption{
\label{fig:timeseries_plot}
(Color online) Normalized, deseasonalized wind generation $R(t)$ (thin black line) and binarized time series (thick black line). The values of the two states $R_0$ and $R_1$ are derived from the full renewables.ninja \cite{Staf16} time series (cf.~\cref{eq:a01_R0,eq:a01_R1}). The load time series is normalized to $L(t) \equiv \langle L \rangle \equiv 1$ (see text) and the wind generation time series is scaled with $\gamma = 1$ such that $\langle R \rangle = 1$ (cf.~\cref{eq:scaling}). Therefore, red shaded areas represent periods of overproduction ($R(t) \geq 1$) and blue shaded areas represent periods of scarcity ($R(t) < 1$). 
}
\end{figure*}

A feature of the operation of the storage system is whether we are in the regime of scarcity ($R(t) < L(t)$) or overproduction ($R(t) > L(t)$). Hence, it is reasonable to consider the wind generation time series as a two-state stochastic system (cf.~\cref{fig:timeseries_plot}, described further in \cref{sec:binarization_of_data}). 		
In this case, we can make use of the relation between the autocovariance function $K(r)$ of the time series and the memory function $F(r)$ (\cref{eq:lin_eq_Fr}) such that it is straightforward to deduce the stochastic model from measured data.

\subsection{Data} \label{sec:Methods_Wind}
We apply the additive binary Markov model to generate synthetic wind generation time series. As input we use the renewables.ninja \cite{Staf16} wind dataset for Germany which consists of hourly wind capacity factors (i.e. wind power normalized by the rated capacity) based on MERRA reanalysis data \cite{Rien11} simulating the 2014 fleet of wind farms for 1985-2014. A wind generation time series exhibits a seasonal as well as a diurnal periodicity \cite{Pesc15,Cara13,Cara13a,Reye16,2017Moemken}: The generation in Germany is usually higher in winter than in summer. The diurnal wind power variation weakly depends on the respective season.
Furthermore, the wind generation time series exhibits `good' and `bad' wind years as well as fluctuations on the synoptic time scale (which is of the order of a few days). Therefore, a wind generation time series is not a stationary process. However, Thomann and Barfield \cite{Thom88} argue that the effect of non-stationarity can be neglected as long as the time series does not exhibit a trend. The latter statement is true because a fixed installed capacity of wind farms was used in order to generate the renewables.ninja \cite{Staf16} data set. Additionally, it is valid to assume that the average wind generation does not exhibit a trend as long as long time frames are considered. As we use 30 years of hourly wind data, this condition is fulfilled.

As our model cannot represent the deterministic, long-range seasonal variations, we seasonally adjust the input data by multiplicatively removing the typical seasonal variation of the wind generation time series using monthly averages. For this purpose, we divide the data of a respective month $R_\text{month}(t)$ by a corresponding seasonality factor $sf_\text{month}$ given as

\begin{equation}
sf_\text{month} = \frac{\langle R_\text{month}(t) \rangle }{\langle R(t) \rangle },
\end{equation}
where $\langle \cdot \rangle$ denotes the average over either all data corresponding to one month or the whole time frame.

We consider different scenarios for the expansion of wind power in the system. Therefore, we scale the capacity factor time series $w(t)$ from renewables.ninja \cite{Staf16} such that the average generation by wind power plants provides a certain share $\gamma$ of the average load, i.e.~\cite{Heid10,Rasm12}
\be
R(t) = \gamma \, \frac{w(t)}{\langle w \rangle} \, \langle L \rangle.
\label{eq:scaling}
\ee

\noindent The factor $\gamma$ is also denoted as renewable penetration. In the case of a fully renewable power system, which we want to consider here primarily, $\gamma$ is set to one. Other scenarios are also discussed (\cref{sec:thr}).

For simplicity, we assume a constant load time series normalized to one ($L(t) \equiv \langle L \rangle \equiv 1$) in the following. In the case of a highly renewable power system, which we consider in this paper, the fluctuations of the wind generation time series are much higher than those of the load time series. Therefore, setting $L$ to a constant value has only a minor influence on backup and storage needs. Furthermore, our primary objective is to evaluate the performance of the additive binary Markov chain model which is facilitated by simplifying the load time series.

\subsection{Binarization of the data} \label{sec:binarization_of_data}

The task of this paper is to develop a simple stochastic model for the intermittent wind time series R(t) -- hence, we want to represent the stochastic hour-by-hour fluctuations, the diurnal cycle and variations on the synoptic time scale. To harness the strength of additive binary Markov chains, we need to define a reasonable mapping from the continuous values of $R(t)$ to a binary time series $a(t)$ with values zero and one. One example for such a mapping is to define a threshold value which separates the wind power generation into regimes of under- ($R(t) < \langle L \rangle$) and overproduction ($R(t) \geq \langle L \rangle$)
\par\nobreak\noindent 
\begin{align}
a(t) = 
\begin{cases}
	0 \quad &{\rm if} \quad R(t) < \langle L \rangle \\
    1 \quad &{\rm if} \quad R(t) \geq \langle L \rangle.
\end{cases}
\label{eq:a01}
\end{align}

From the binarized time series $a(t)$ we can derive the autocovariance function $K(r)$ (cf.~\cref{eq:Kr}). Choosing an appropriate memory length $N$, the memory function $F(r)$ can be recovered using \cref{eq:lin_eq_Fr} and the Markov chain can be simulated using \cref{eq:add_markov}. 

In order to apply the power balance equation (\cref{eq:power_balance}) to the binarized time series, we map $a(t)$ to average values of under- and overproduction of the input time series (see \cref{fig:timeseries_plot}),
\par\nobreak\noindent 
\begin{align}
a(t) = 0 \quad \Rightarrow \quad R_{\rm Bin}(t) = R_0 &= \langle R \rangle |_{R(t) < \langle L \rangle}, \label{eq:a01_R0} \\ 
a(t) = 1 \quad \Rightarrow \quad R_{\rm Bin}(t) = R_1 &= \langle R \rangle |_{R(t) \geq \langle L \rangle},
\label{eq:a01_R1}
\end{align}
\noindent where $\langle \cdot \rangle |_x $ is the conditional expectation value. As we want to fix the renewable penetration $\gamma$ for all simulations, the simulated values for $R_0$ and $R_1$ have to be scaled further such that $\langle R \rangle = \gamma$ always holds (cf.~\cref{eq:scaling}).

\subsection{Initialization of the model} \label{sec:initialization}

The synthetic time series are modeled in such a way that the original and the simulated time series are of the same length (30 years in hourly resolution). In each case, an ensemble of 100 Monte Carlo simulations is considered.

The first $N$ elements in the synthetic sequence are generated using the following modification of \cref{eq:add_markov}:
\par\nobreak\noindent
\begin{align}
P(a_1=1) &=	\langle a \rangle \\
\text{if} \quad t &= 1 \nonumber \\
P(a_t=1 | T_t) & = \langle a \rangle + \sum\limits_{r=1}^t F(r) \, (a_{t-r} - \langle a \rangle) \\
\text{if} \quad 1 &< t \leq N. \nonumber
\end{align}
\noindent This is considered to be a reasonable assumption, as the first $N$ values represent only a very small fraction of the entire simulated stochastic sequence.

\section{Results} \label{sec:Results}
In the following sections, we discuss the memory function of the binarized time series used in this paper (\cref{sec:Fr}) and subsequently evaluate our additive binary Markov chain model by analyzing the autocorrelation function, the backup and storage need and the resting time distribution for different memory lengths $N \in \{7, 14, 21, 70\} \times 24$ hours in comparison to a simple, memoryless Markov chain (\cref{sec:ABMC}). Additionally, we assess the impact of the threshold value used to binarize the input data (\cref{sec:thr}) by choosing different values for the renewable penetration $\gamma$.

\subsection{Memory function} \label{sec:Fr}
\begin{figure}[tb]
\centering
\includegraphics[width=0.48\textwidth]{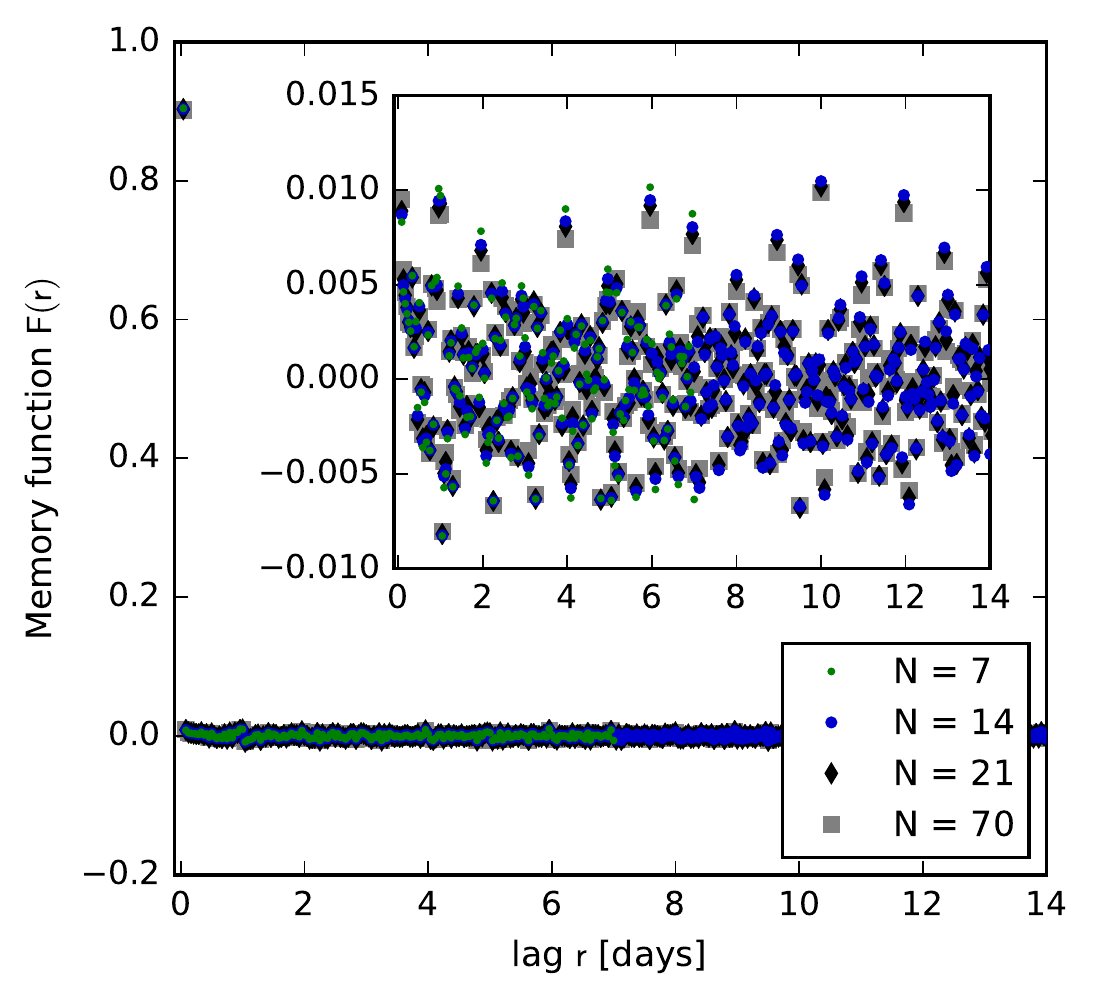}
\caption{
\label{fig:F_r}
(Color online) Memory function $F(r)$ for memory lengths of $N \in \{7, 14, 21, 70\}$ days (i.e.~green dots, blue circles, black diamonds and grey boxes, respectively). The inset shows a zoom into the region around $F(r) = 0$ for better visualization. The time lag $r$ is given in hourly resolution.
}
\end{figure}

In order to simulate binary wind generation data using the additive binary Markov model, a memory function $F(r)$ has to be derived by solving the set of linear equations in \cref{eq:lin_eq_Fr} for a predefined memory length $N$. In \cref{fig:F_r} the memory function $F(r)$ is shown for memory lengths $N \in \{7, 14, 21, 70\} \times 24$ hours. The memory function is hardly sensitive to small changes of $N$. For all four choices of $N$, it takes a value of 0.90 for $r=1$ hour and is close to zero for $r>1$ hour (shown in detail in the inset in \cref{fig:F_r}). The inset further shows that $F(r)$ decreases monotonically up to $r = 7$ hours (and non-monotonically up to $r \approx 12$ hours) and exhibits peaks corresponding to a diurnal circle, according to the diurnal variation of the input time series.

The high difference in the magnitude of $F(1)$ and $F(r>1)$ imposes the question if it is possible to obtain the same results when only using $F(1)$ for the simulation. This would imply that the stochastic process is memoryless and can be described by a simple 1-step binary Markov chain. To evaluate this, we also simulate data using only $F(1)$ in \cref{eq:add_markov}, setting $F(r>1) = 0$. In the following, we thus compare the results of the additive binary Markov model with memory to (i) the empirical binary time series and (ii) the results of the binary Markov chain without memory.

\subsection{Additive Markov chain of the binarized wind power data} \label{sec:ABMC}

\subsubsection*{Autocorrelation function}

\begin{figure*}[tb]
\centering
\includegraphics[width=0.99\textwidth]{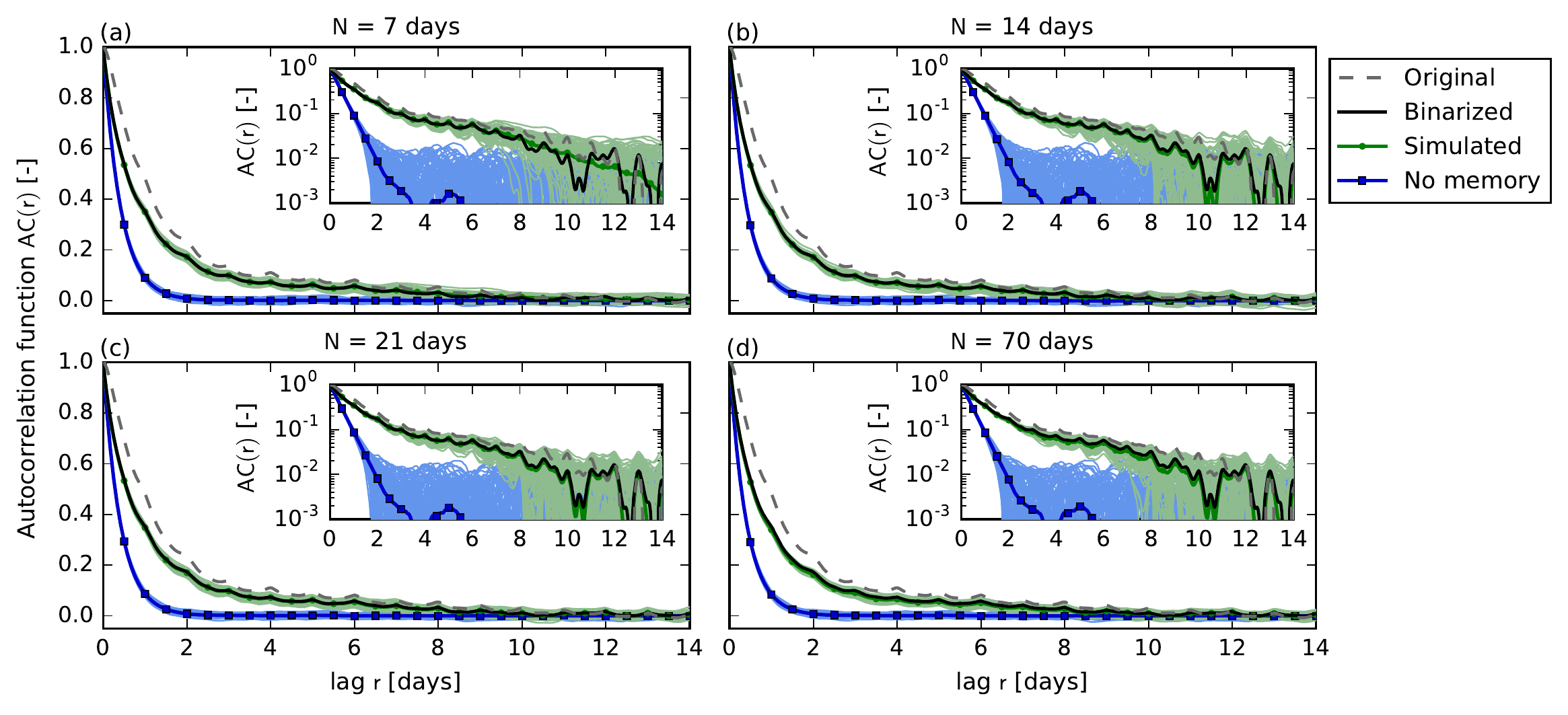}
\caption{
\label{fig:autocorr}
(Color online) Autocorrelation functions of the original renewables.ninja \cite{Staf16} (grey dashed) and the binarized (black) time series and of an ensemble of 100 simulations using the additive binary Markov chain approach for memory lengths of $N \in \{7, 14, 21, 70\}$ days (green dots) and without memory (i.e.~$F(r>1) = 0$, blue squares). The ensemble averages are represented as thick green and blue lines, respectively. For better visualization of the long-time decay, the autocorrelation functions are also shown on a logarithmic y-scale in the insets.
}
\end{figure*}

The autocorrelation function $AC(r)$ -- i.e.~the autocovariance function $K(r)$ normalized such that $AC(0) = 1$ -- describes how strong the wind generation at one time step is correlated to the wind generation at a different, lagged time step ($r>0$). We observe that for the original as well as for the binarized time series, the autocorrelation function decreases fast for time lags up to about $r=3$ days and levels off for higher time lags, being close to zero for about $r > 10$ days (grey dashed and black lines in \cref{fig:autocorr}). Additionally, the curves are superimposed by a slight 24-hour oscillation, representing the diurnal variation of the time series. The binarization of the original time series leads to an underestimation of the autocorrelation function for all lags $r$. This can be explained in the following way: In the binarized time series, the wind generation can only take two values, zero and one. Therefore, the absolute value of the correlation (cf.~\cref{eq:Kr}) is either high (i.e.~$a(t+r) = a(t)$) or low (i.e.~$a(t+r) \neq a(t)$). Small changes are neglected. Because of this, the autocorrelation function of the binarized time series, $AC_{\rm Bin}$, must be used as a reference in order to evaluate the additive binary Markov chain method.

The ensemble-average of the autocorrelation functions derived from the additive binary Markov chain with memory lengths $N \in \{7, 14, 21, 70\}$ days ($AC_\text{Markov}$, thick green line in \cref{fig:autocorr}) is almost identical to the empirical $AC_\text{Bin}$ for all lags $r$ and all memory lengths $N$. Even the diurnal oscillation can be recovered. The autocorrelation function can only be reconstructed for time lags up to $N$ as the conditional probability $P(a_t = 1 | T_{N,t})$ in \cref{eq:add_markov} directly depends on $N$. For higher time lags, the fluctuations in the autocorrelation function cannot be reproduced any longer (cf.~\cref{fig:autocorr}(a)). Furthermore, for high memory lengths (i.e.~$N > 21$ days), the autocorrelation function tends to be slightly underestimated for all values of the time lag $r$. Hence, the memory length $N$ must be chosen long enough to capture all significant correlations in the original binarized data, but not longer. Indeed, choosing a medium value of about $N = 7 \dots 21$ days, the additive binary Markov model reproduces the correlation structure of the binarized wind generation time series almost perfectly.

Without memory (i.e.~$F(r>1) = 0$, blue lines in \cref{fig:autocorr}), the autocorrelation function is strongly underestimated. The correlation approaches zero for time lags of $r > 2$ days. Hence, without memory, the long-range correlations of the wind generation time series cannot be represented. We conclude that even though the values of $F(r)$ are small for $r > 1$ hour (cf. \cref{fig:F_r}), they still contain important information for the simulation.

\subsubsection*{Backup and storage need}

\begin{figure*}[tb]
\centering
\includegraphics[width=0.99\textwidth]{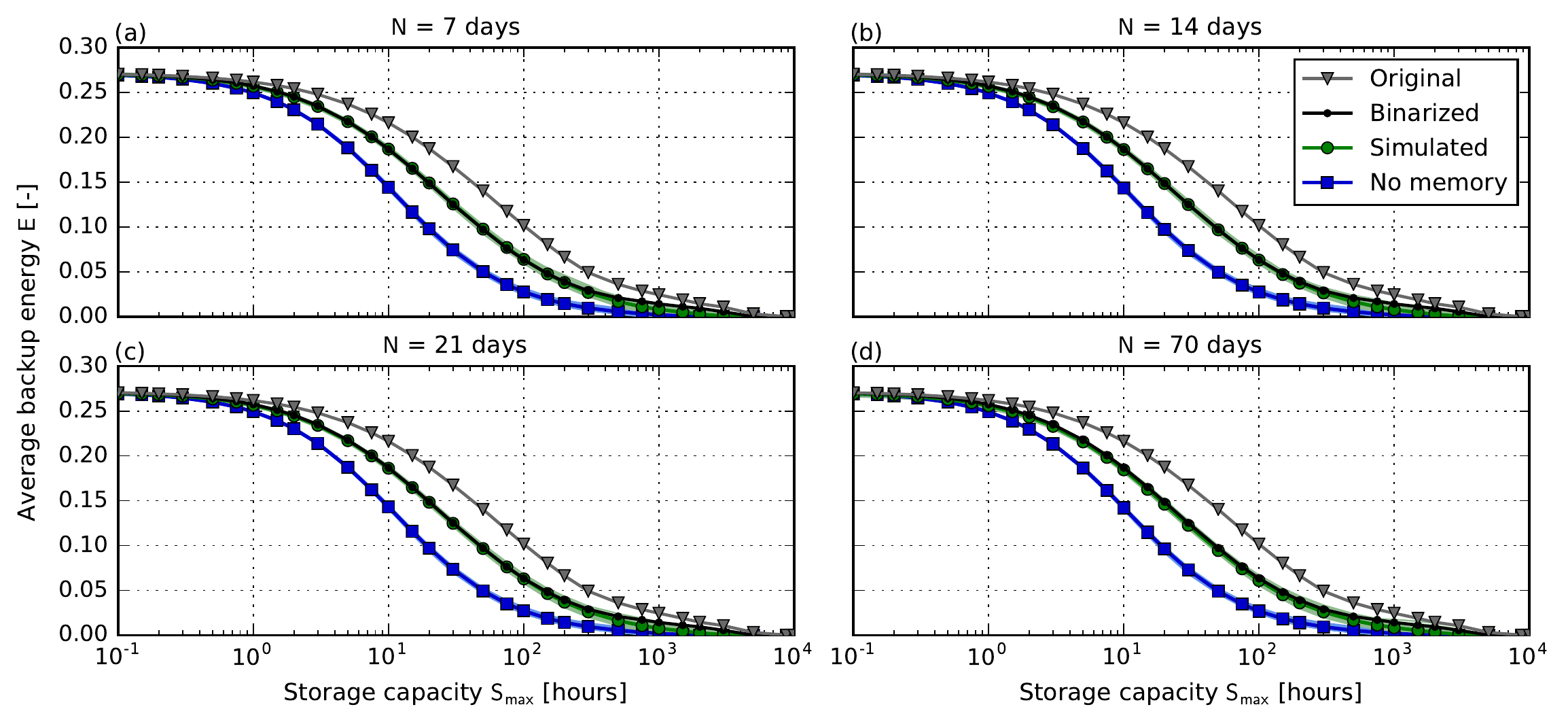}
\caption{
\label{fig:E_Smax}
(Color online) Average backup energy $E$ as a function of the storage capacity $S_\text{max}$ for the original renewables.ninja \cite{Staf16} (grey triangles) and the binarized (black dots) time series and for an ensemble of 100 simulations using the additive binary Markov chain approach (green circles) for memory lengths of $N \in \{7, 14, 21, 70\}$ days and without memory (i.e.~$F(r>1) = 0$, blue squares). The ensemble averages are represented as thick green and blue lines.
}
\end{figure*}

A reliable wind power model should reproduce the essential characteristics of the original (binarized) time series. A key benchmark for power system operation is the average amount of backup energy $E$ as a function of the storage capacity $S_{\rm max}$. In order to keep the assumption of a fully-renewable power system, the simulated data is scaled such that the relation $\langle R \rangle = \langle L \rangle = 1$ always holds.

The mapping of the original data to binary values reduces the backup need for all given $S_{\rm max}$ (grey triangles and black dots lines in \cref{fig:E_Smax}) due to the loss of temporal correlation as discussed above. Thus, in order to evaluate our method, we compare the average backup energy of the simulated time series to that derived from the binarized time series ($E_{\rm Bin}$).

Considering the additive binary Markov chain model with memory (green lines in \cref{fig:E_Smax}), we find that the average backup energy need, $E_\text{Bin}(S_\text{max})$, is reproduced with high accuracy for most storage capacities $S_\text{max}$. For $S_\text{max} > 100$ hours, the backup energy is underestimated a little for all $N$. For high memory lengths (i.e., $N = 70$ days, \cref{fig:E_Smax}(d)), the backup energy is slightly underestimated for all storage capacities $S_\text{max}$. This can be explained by the overall underestimation of the temporal correlation (cf.~\cref{fig:autocorr}(d)).

The additive binary Markov chain without memory (i.e.~$F(r>1)=0$) highly underestimates the backup need (blue lines in \cref{fig:E_Smax}). This results from (i) the underestimation of long-range correlations (cf.~\cref{fig:autocorr}) and (ii) the fact that the resting time distribution of under- and overproduction events cannot be reconstructed (see \cref{fig:duration} below). This shows that the memory function $F(r)$ has to be considered for $r>1$ hour in order to get reasonable results for practically relevant characteristics of the time series.

\subsubsection*{Resting time distribution}

\begin{figure*}[tb]
\centering
\includegraphics[width=0.99\textwidth]{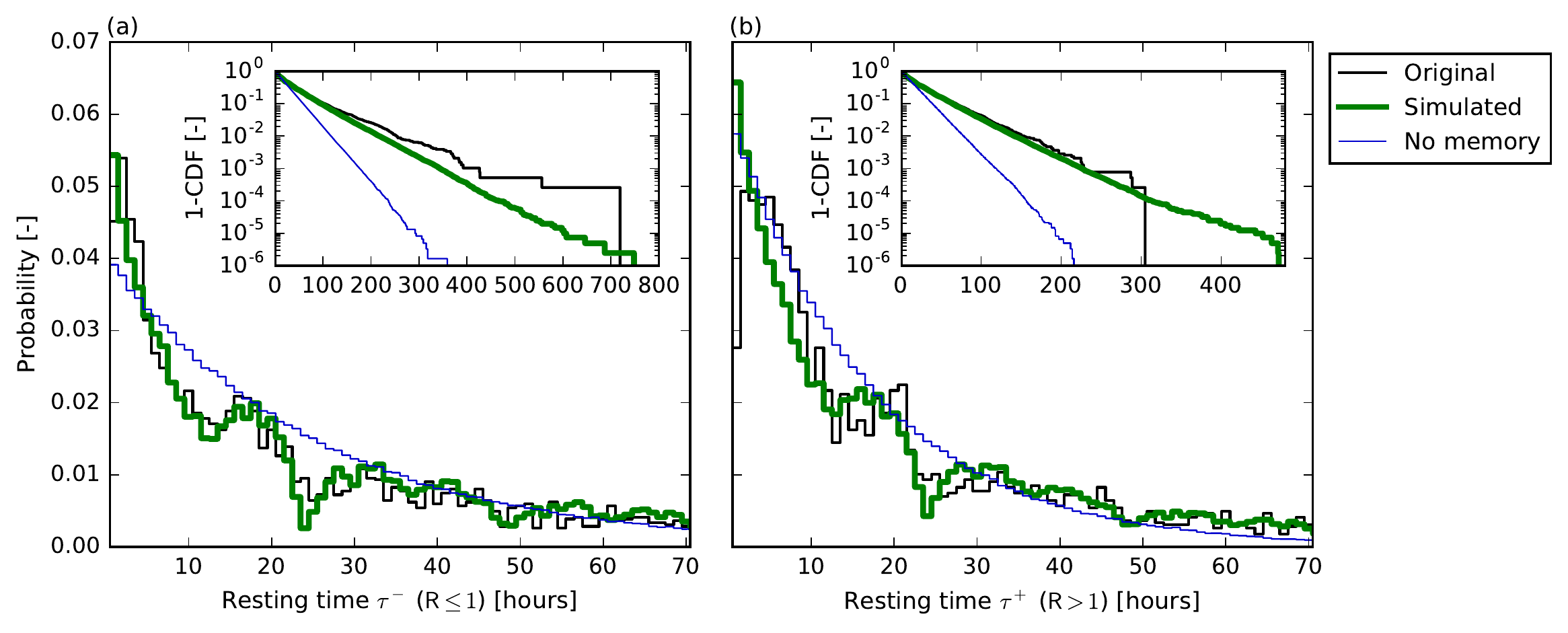}
\caption{
\label{fig:duration}
(Color online) Resting time distribution of the original renewables.ninja \cite{Staf16} time series (medium black lines) and the ensemble of 100 simulations using the additive binary Markov chain approach for a memory length of $N=14$ days (thick green lines) and without memory (i.e.~$F(r>1) = 0$, thin blue lines). Panel (a) shows the resting times for which the renewable generation $R$ is constantly smaller than the (constant) load $L=1$ (underproduction) and panel (b) shows the resting times for overproduction. The main figures show the probability of resting times up to 70 hours whereas the insets display the probability of long resting times using one minus the cumulative distribution function (CDF) and a logarithmic y-scale.
}
\end{figure*}

A further benchmark for the performance of the additive binary Markov approach is to evaluate the distribution of the duration $\tau$ of periods for which $a(t) = 0$ (denoted as $\tau^-$) and $a(t) = 1$ (denoted as $\tau^+$), respectively (cf.~\cref{eq:a01_R0,eq:a01_R1}). A proper representation of this resting time distribution is particularly important for the backup and storage needs (cf.~\cref{fig:E_Smax}). This is especially true for the case of long lasting scarcity events (i.e.~$R(t) < 1$), which can deplete the storage and lead to high needs of backup energy. The resting time distribution of the synthetic data is derived from the ensemble of 100 simulations. Results are shown for $N=14$ days in \cref{fig:duration}. We find that they hardly depend on the choice of $N$. 

The distributions of the original and the simulated time series with memory are well comparable for a wide range of resting times (black and green lines in \cref{fig:duration}). Resting  times for underproduction (i.e.~$R \leq 1$) are slightly underestimated for $1 < \tau^- \leq 4$ hours. Furthermore, resting times for overproduction (i.e.~$R > 1$) tend to be overestimated for $\tau^+ \leq 3$ hours and are underestimated a little for $3 < \tau^+ \leq 9$ hours. This effect may result from the daily wind variation of the data. This leads to a maximum in the original resting time distribution at $\tau^\pm = 2$ hours. This maximum cannot be represented by our model and is rather smoothed out. 

Due to the diurnal variation in the wind power generation, we find local minima for resting times of about 12 and 24 hours for both, under- and overproduction. These local minima are well represented by our model.

The probability for a resting time $\tau^\pm$ to be longer than a certain value(shown as one minus the cumulative distribution function in the insets in \cref{fig:duration}) decreases almost exponentially for the simulated data with memory. In contrast, for the original data the probabilities for long resting times decay more slowly. This most likely results from the small amount of original data showing high resting times and hence represents a finite-size effect. In the original dataset, no resting times longer than $\tau_\text{max}^- = 719$ hours and $\tau_\text{max}^+ = 305$ hours are present such that it is impossible to compare probabilities in this regime.

Without memory (i.e.~ $F(r>1)=0$, blue lines in \cref{fig:duration}), the resting time distribution cannot be reconstructed properly. The curve also decreases exponentially, however, with the wrong rate. Especially, the probability for long durations is highly underestimated. This again shows that the higher terms of the memory function are essential for the stochastic model to work.

\subsubsection*{Conclusion}
The additive binary Markov chain model reproduces the temporal characteristics of the binarized wind power time series to a high extent when using the full memory function $F(r)$. The autocorrelation function is reproduced almost perfectly within the memory time per construction, provided the memory time is not chosen unreasonably high. The backup and storage needs can be represented very well -- especially for small and medium storage sizes. Finally, the resting time distribution of periods with below- and above-threshold wind generation can be reconstructed. Deviations are found for very large time spans, but the comparison to the original time series becomes tedious here due to finite-size effects. In comparison, an additive binary Markov model without memory clearly fails to reproduce characteristics essential for the power system operation and design. This proves that long memories must be taken into account in any stochastic model for wind power generation. The memory length $N$ should be of the same order of magnitude as the time scales of the temporal correlations considered in the simulation (here: $N = 7 \dots 21$ days).

\subsection{Performance for different values of the renewable penetration} \label{sec:thr}
So far, we assumed a fully renewable power system with 100\% renewable generation on average (i.e.~$\gamma = 1$). We now evaluate the model for different scenarios for the development of renewable power sources quantified by the renewable penetration $\gamma$. This corresponds to different threshold values in the binarization of the empirical time series (cf.~\cref{eq:a01}). We scale the wind power time series using (i) $\gamma = 0.5$ and (ii) $\gamma = 1.5$ (cf.~\cref{eq:scaling}) such that the average wind generation equals (i) 50\% and (ii) 150\% of the load, indicating long resting times for under- and overproduction, respectively. As before, we consider an ensemble of 100 simulations and choose a memory length of $N = 14$ days.

\subsubsection*{Autocorrelation function}

\begin{figure*}[tb]
\centering
\includegraphics[width=0.99\textwidth]{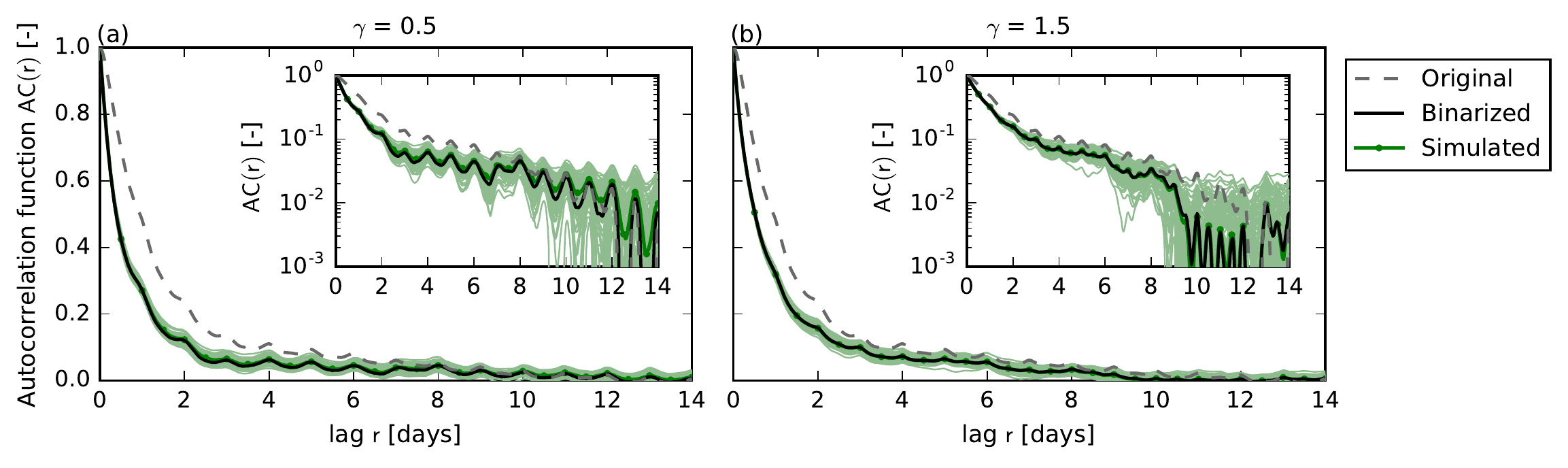}
\caption{
\label{fig:gamma_autocorr}
(Color online) Autocorrelation functions for the original renewables.ninja \cite{Staf16} (grey dashed) and the binarized (black) time series and an ensemble of 100 simulated time series using the additive binary Markov chain approach (green dots) for renewable penetrations of $\gamma = 0.5$ (panel (a)) and $\gamma = 1.5$ (panel (b)) and a memory length of $N=14$ days. The ensemble average is represented as thick green line.
}
\end{figure*}

For both values of $\gamma$, the ensemble average of the autocorrelation function derived using the additive binary Markov model coincides almost perfectly with the autocorrelation function of the original binarized time series (see \cref{fig:gamma_autocorr}). Hence, we conclude that the scaling, and thus the choice of the threshold value, does not impact the representation of the temporal correlations in our model.

\subsubsection*{Backup and storage need}

\begin{figure*}[tb]
\centering
\includegraphics[width=0.99\textwidth]{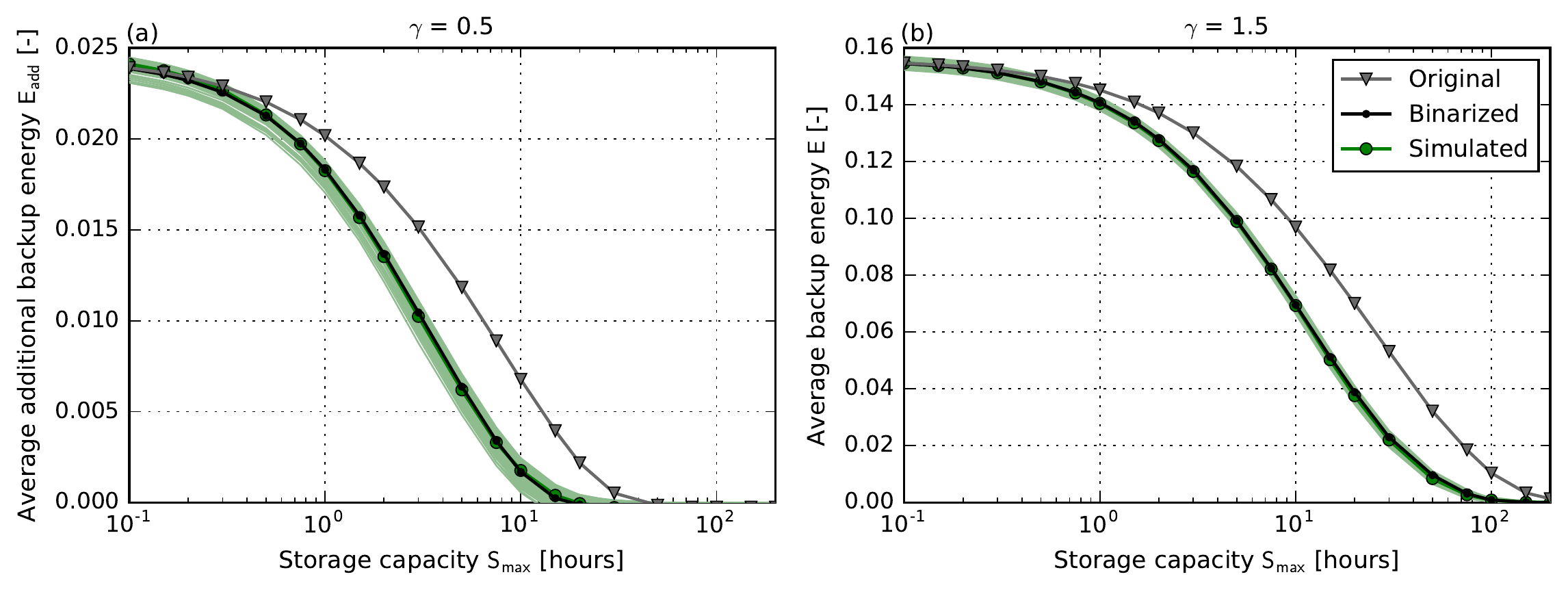}
\caption{
\label{fig:gamma_E_Smax}
(Color online) Average (additional) backup need $E$ ($E_\text{add}$) as a function of the storage capacity $S_\text{max}$ for the original renewables.ninja \cite{Staf16} (grey triangles) and the binarized (black dots) time series and an ensemble of 100 simulated time series using the additive binary Markov chain approach (green circles) for renewable penetrations of $\gamma = 0.5$ (panel (a)) and $\gamma = 1.5$ (panel (b)) and a memory length of $N=14$ days. The ensemble average is represented as thick green line.
}
\end{figure*}

If the renewable penetration is chosen to be smaller than one, an average share of $1-\gamma$ has to be provided by non-renewable power plants. This share is deterministic. We are interested in the non-deterministic additional backup energy which is given as $E_\text{add} = E -(1-\gamma)$. 

The (additional) average backup energy as a function of the storage capacity is shown in \cref{fig:gamma_E_Smax}. A comparison with \cref{fig:E_Smax} shows that for $\gamma \neq 1$, the (additional) average backup energy is smaller than for $\gamma = 1$ for all storage capacities. This observation is well analyzed in the literature \cite{Heid10,Rasm12,Jens14}. The average backup energy need can be reproduced almost perfectly for both, a scaling with $\gamma = 0.5$ and a scaling with $\gamma = 1.5$. For $\gamma = 0.5$ and $S_\text{max} < 0.3$ hours, $E_\text{add}$ is slightly overestimated whereas it is underestimated a little bit for $ 1.5 \leq S_\text{max} \leq 7.5$ hours. This may be explained by the small differences in the resting time distribution (see \cref{fig:gamma_duration} in the following section). For $\gamma = 1.5$, $E$ is also slightly underestimated for $ 15 \leq S_\text{max} \leq 75$ hours which may be due to an underestimation of the probability for resting times $\tau^+$ to be longer than about 200 hours.

\subsubsection*{Resting time distribution}

\begin{figure*}[tb]
\centering
\includegraphics[width=0.99\textwidth]{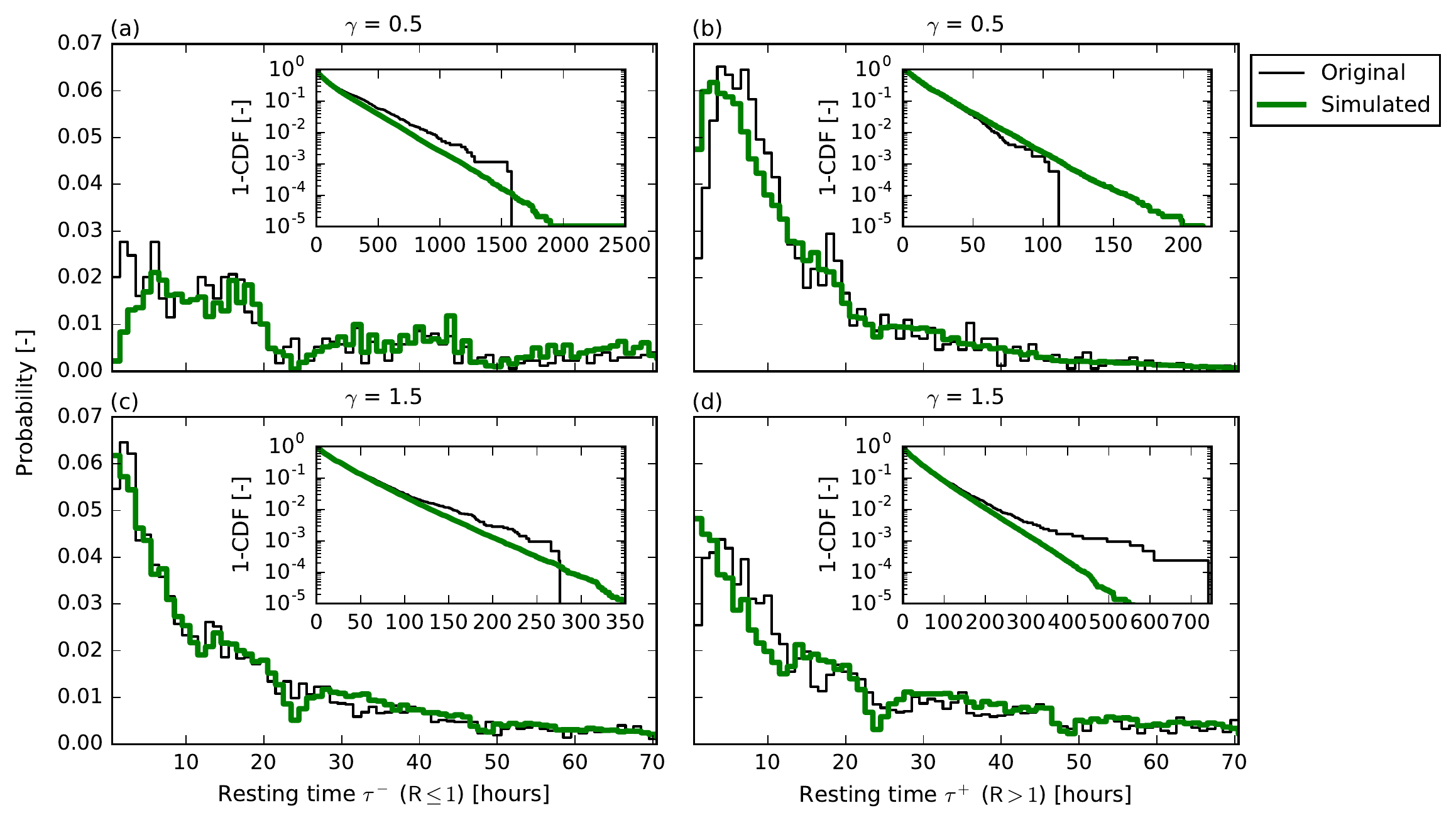}
\caption{
\label{fig:gamma_duration}
(Color online) Resting time distribution of the original renewables.ninja \cite{Staf16} time series (thin black lines) and the ensemble of 100 simulations using the additive binary Markov chain approach (thick green lines) for a memory length of $N=14$ days. Panels (a) and (c) show the resting times for underproduction ($R \leq 1$) and panels (b) and (d) show the resting times for overproduction ($R > 1$) for renewable penetrations of $\gamma = 0.5$ (top row) and $\gamma = 1.5$ (bottom row). The main figures show the probability of resting times up to 70 hours whereas the insets display the probability of long resting times using one minus the cumulative distribution function (CDF) and a logarithmic y-scale.
}
\end{figure*}

For $\gamma=0.5$, the resting time distribution exhibits longer resting times $\tau^-$ for periods of underproduction and shorter resting times $\tau^+$ for periods of overproduction (see \cref{fig:gamma_duration}(a) and (b)) compared to $\gamma = 1$ (cf.~\cref{fig:duration}), per construction. In the case of $\gamma = 1.5$ the opposite is true (see \cref{fig:gamma_duration}(c) and (d)).

For resting times up to 70 hours, the resting time distributions are represented well in most cases. For $\gamma = 0.5$, the model tends to underestimate the probability for $\tau^- \leq 6$ hours (\cref{fig:gamma_duration}(a)). 
Furthermore, as in the case of $\gamma = 1$, the model cannot reproduce the local maximum at $\tau^\pm \approx 2-4$ hours very well. This is especially true for the case of overproduction (i.e., $R > 1$).

Considering the probability for resting times being longer than a certain value (insets in (\cref{fig:gamma_duration}), the resting time distributions of the original and the simulated data are well comparable and the slopes are almost identical. Due to the large ensemble of 100 simulations, the simulated data extends to longer resting times than the original data. For $\gamma = 1.5$ and $\tau^+$, the probability for long durations is lower in the simulations than in the original time series.

\subsubsection*{Conclusion}
In conclusion, the additive binary Markov model well reconstructs the temporal correlations and the backup energy as a function of the storage capacity for different scenarios of the development of the renewable expansion $\gamma$ (i.e.~for different threshold values). Furthermore, our model is also capable to represent very long resting times, especially in the case of underproduction which is important for the sizing of the backup and storage infrastructure.

\section{Extension of the model to non-binary wind power data} \label{sec:tauA}

One major drawback of our model is that it relies on binarized data, which underestimates the autocorrelation function as well as the average backup need $E$. We stress that this is not due to the stochastic modeling but solely due to the simplification of the empirical input data. We now present a method to overcome this deficiency, while keeping the simplicity of the binary model.

\subsection{Relation between duration and energy}

\begin{figure*}[tb]
\centering
\includegraphics[width=0.99\textwidth]{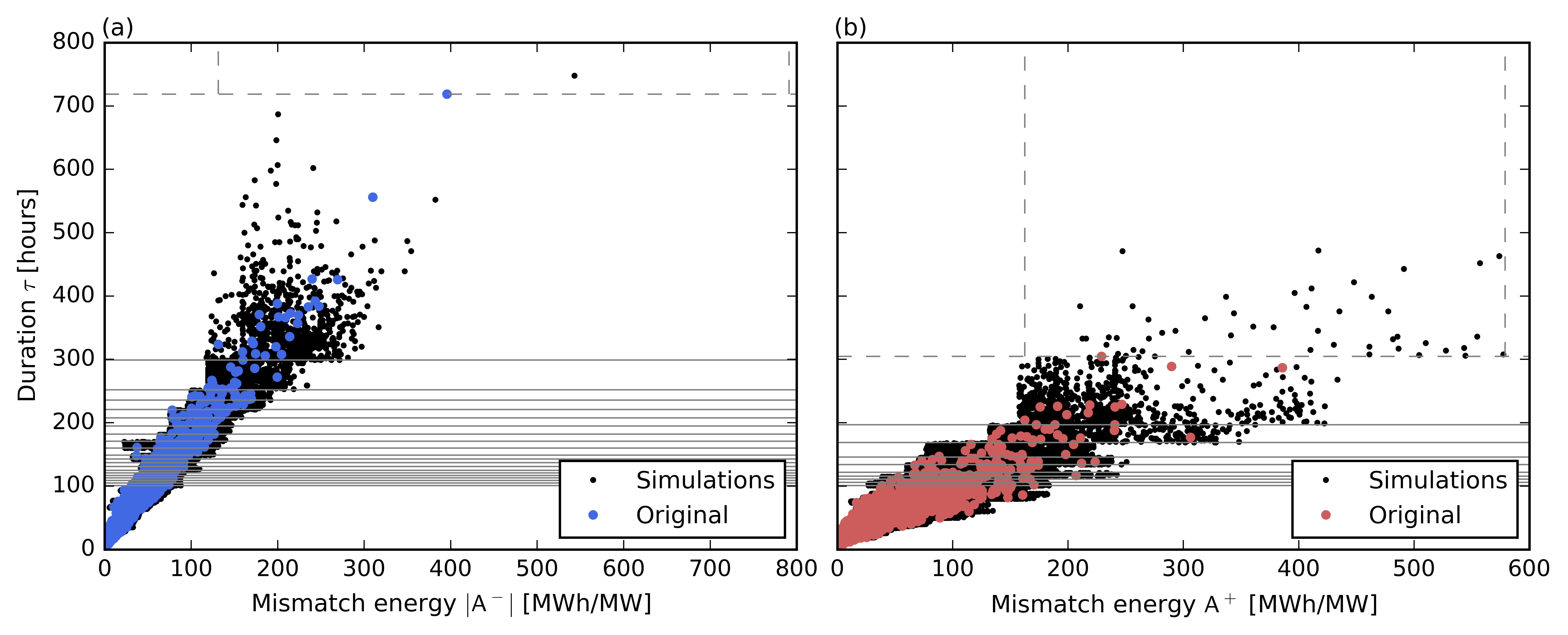}
\caption{
\label{fig:A-tau}
(Color online) Scatter plot of the joint distribution of the duration $\tau$ and the corresponding mismatch energy of under- ($|A^-|$, panel (a)) and overproduction ($A^+$, panel (b)) for the original renewables.ninja \cite{Staf16} time series (large blue and red dots) and for an ensemble of 100 simulations using the additive binary Markov chain approach (small black dots) for a memory length of $N=14$ days. The $\tau$-bins are chosen such that at least 15 of the original values lie within one bin (see text). They are represented by the horizontal grey lines and only shown for $\tau \geq 100$ hours. For each simulated $\tau$, an energy value $A$ is chosen according to the distribution of the original values within the bin. Energies which lie outside the range of the original values result from a rescaling as described in detail in the main text. If the simulated duration is higher than the maximum duration in the original time series (horizontal grey dashed line), the energy values are distributed uniformly within the region represented by the vertical grey dashed lines.
}
\end{figure*}

The deterministic mapping (\cref{eq:a01_R0,eq:a01_R1}) is replaced by a stochastic mapping
\par\nobreak\noindent 
\begin{align}
a(t) = 0 \quad \Rightarrow \quad R(t) &= R^- \label{eq:a01_Rsim0} \\ 
a(t) = 1 \quad \Rightarrow \quad R(t) &= R^+, \label{eq:a01_Rsim1}
\end{align}
\noindent where $R^-$ and $R^+$ are random variables. They remain constant throughout any period of scarcity or oversupply, respectively, but depend on the duration of the period $\tau^\pm$.

The stochastic mapping is constructed from the empirical time series as follows: In each period of scarcity or oversupply, a certain amount of energy has to be provided by the backup and/or storage facilities, or curtailed and/or stored in, respectively. The fundamental variable is thus the mismatch energy of one period. We use this energy as the basic variable for the stochastic mapping and derive it as the area ($A^-$ and $A^+$) under the curve for which $R(t) - \langle L \rangle < 0$ or $R(t) - \langle L \rangle  > 0$, respectively (see \cref{fig:timeseries_plot}):
\be
A^\pm(\tau^\pm) = \int_{t_1}^{t_2} [ R(t)- \langle L \rangle ] \, \mathrm{d}t
\ee
\noindent with $\tau^\pm = t_2 - t_1$ and $\langle L \rangle = 1$. 

A strong positive correlation exists between the duration $\tau^\pm$ of such a period and the corresponding mismatch energy $A^\pm$ (see \cref{fig:A-tau}, colored dots): The longer the duration of an event, the higher the mismatch energy, on average. We use this observation to assign mismatch energies to the periods of over- and underproduction simulated using the additive binary Markov chain. In order to do this, we derive a relation between $\tau^\pm$ and $A^\pm$ from the original renewables.ninja \cite{Staf16} dataset: First, we define bins for $\tau^+$ and $\tau^-$ such that at least 15 values lie within each bin (cf.~\cref{fig:A-tau}, horizontal lines). $\tau$ is a discrete variable. Thus, for short periods of over- and underproduction, for which more than 15 observations exist, the binwidth is set to one hour, the time resolution of the input data. The last bin is chosen such that it contains at least 10 observations. Second, we calculate the empirical cumulative distribution function (ECDF) of the corresponding energies $A^+$ and $A^-$ for every $\tau^\pm$-bin. Using a linear interpolation of the inverse of the ECDF, we can randomly assign energies $A^\pm$ for each duration $\tau^\pm$ encountered in the additive binary Markov model. If the simulated duration is higher than any observed duration in the original time series, the mismatch energy is chosen randomly within an empirically chosen interval (vertical dashed lines in \cref{fig:A-tau}).

In a fully-renewable power system (i.e.~$\gamma=1$) the sum of all mismatch energies has to be zero, i.e.~$\sum A^+ + \sum A^- = 0$. In the simulations, this sum is usually only close to zero. Thus, the energies have to be rescaled. If $\sum A^+ < \sum |A^-|$, the $A^+$ are scaled to higher values:
\be
A^+_{\rm new} = \frac{\sum |A^-|}{\sum A^+} \cdot A^+
\label{eq:A_scaling}
\ee
and vice versa. It should be noted that this scaling can be done in several ways. We chose to scale to higher absolute values, because we find that the ensemble mean of the average mismatch energy ($\frac{1}{100} \cdot \sum_{n=1}^{100} \langle A^\pm \rangle$) is smaller than the average mismatch energy of the original time series.

The black dots in \cref{fig:A-tau} resemble the resulting $\tau$-$A$ distribution of the ensemble of 100 simulations of the additive Markov chain for $N$=14 days and $\gamma = 1$. 

\subsection{Evaluation of the method}

\begin{figure*}[tb]
\centering
\includegraphics[width=0.99\textwidth]{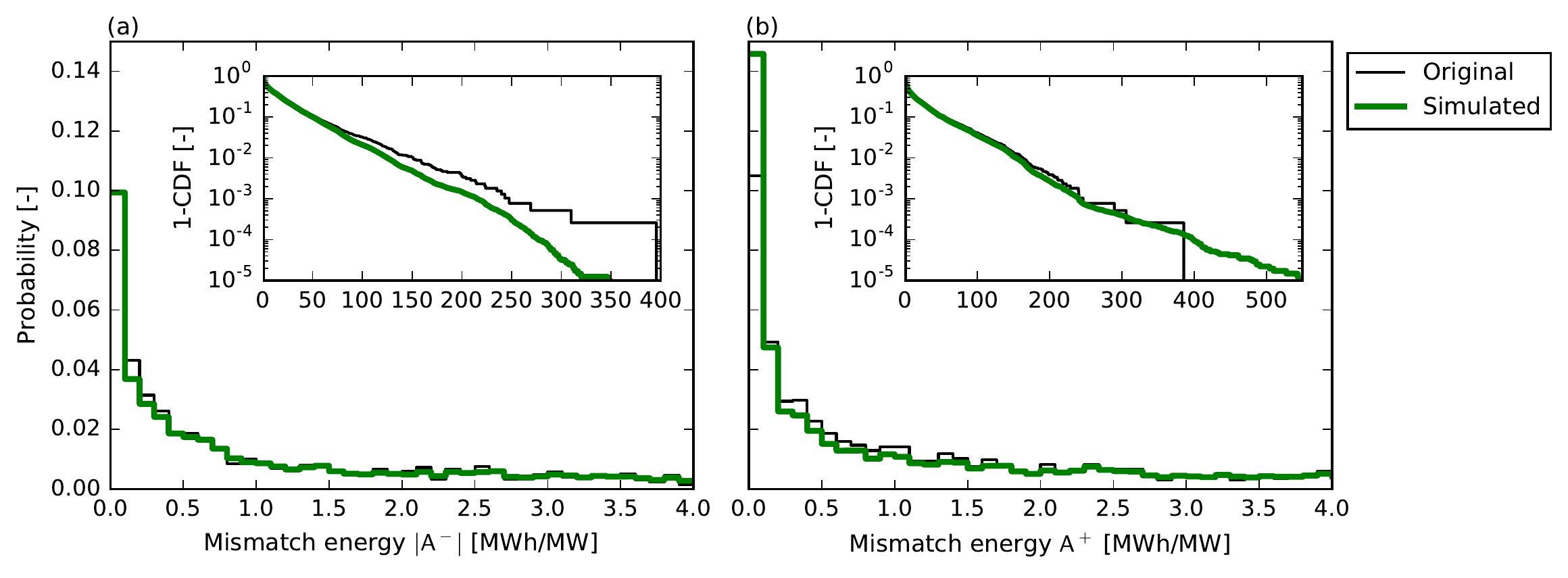}
\caption{
\label{fig:A_marginal}
(Color online) Marginal distribution of the energies corresponding to under- ($|A^-|$, panel (a)) and overproduction ($A^+$, panel (b)) for the original renewables.ninja \cite{Staf16} time series (thin black lines) and the ensemble of 100 simulations using the additive binary Markov chain approach for a memory length of $N=14$ days (thick green lines). The main figures show the probabilities for small amounts of mismatch energies whereas the insets focus on large mismatch energies using one minus the cumulative distribution function (CDF) and a logarithmic y-scale.
}
\end{figure*}

\subsubsection*{Marginal distributions of the mismatch energies}

The marginal distributions of $A^+$ and $A^-$ are almost identical for the original and the simulated time series for a wide range of values (\cref{fig:A_marginal}). 
The simulations slightly underestimate the probabilities for $ 0.1 < |A^-| < 0.5$ and $ 0.1 < A^+ < 1.2$. This results directly from the deviations in the resting time distribution (cf.~\cref{fig:duration}): As $\tau$ and $A$ are strongly correlated, an underestimated probability for small values of $\tau$ most certainly leads to an underestimated probability for small mismatch energies as well.
For high absolute values of $A^+$ and $A^-$ (insets of \cref{fig:A_marginal}), the same arguments hold: The probabilities for $|A^-| > 50$ and $150 < A^+ < 386$ are smaller in the simulations compared to the original time series due to the lower likelihood for durations in the range of $100 < \tau^- < 719$ and $100 < \tau^+ < 305$, respectively which results from finite-size effects.

\begin{figure}[tb]
\centering
\includegraphics[width=0.48\textwidth]{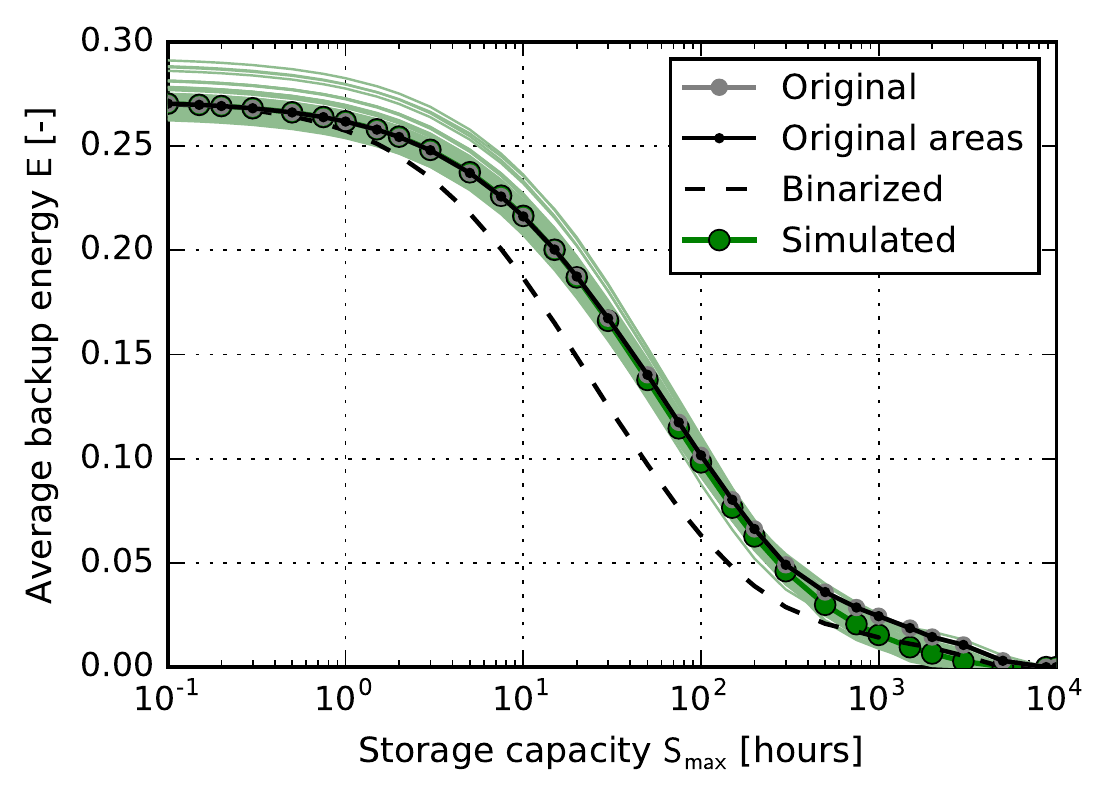}
\caption{
\label{fig:E_Smax_delta}
(Color online) Average backup energy $E$ as a function of the storage capacity $S_\text{max}$ for the ensemble of 100 simulations using the additive binary Markov chain approach combined with the mismatch energy distribution (green circles) for a memory length of $N = 14$ days. The ensemble average is represented as thick green line. Results have to be compared to those of the piece-wise averaged original renewables.ninja \cite{Staf16} time series (black dots). Additionally, results for the original time series (grey circles) and for the binarized time series (black dashes) are shown.  
}
\end{figure}

\subsubsection*{Backup and storage need}

To test our model, we again derive the average backup energy $E$ as a function of the storage capacity $S_{\rm max}$. In order to apply \cref{eq:power_balance,eq:storage}, we calculate the piece-wise averaged wind generation time series from the mismatch energies as follows:
\be
R^\pm(t) = \langle L \rangle + \frac{A^\pm(\tau^\pm)}{\tau^\pm}.
\ee

The average over the ensemble of 100 simulations almost perfectly coincides with the average backup energy $E$ of the original time series for $S_\text{max} < 50$ hours (see \cref{fig:E_Smax_delta}). For higher storage capacities, $E$ is slightly underestimated as already observed before (cf.~\cref{fig:E_Smax}(b)). The spread of the ensemble however is higher because the mismatch energies are not perfectly matched.

\begin{figure}[tb]
\centering
\includegraphics[width=0.48\textwidth]{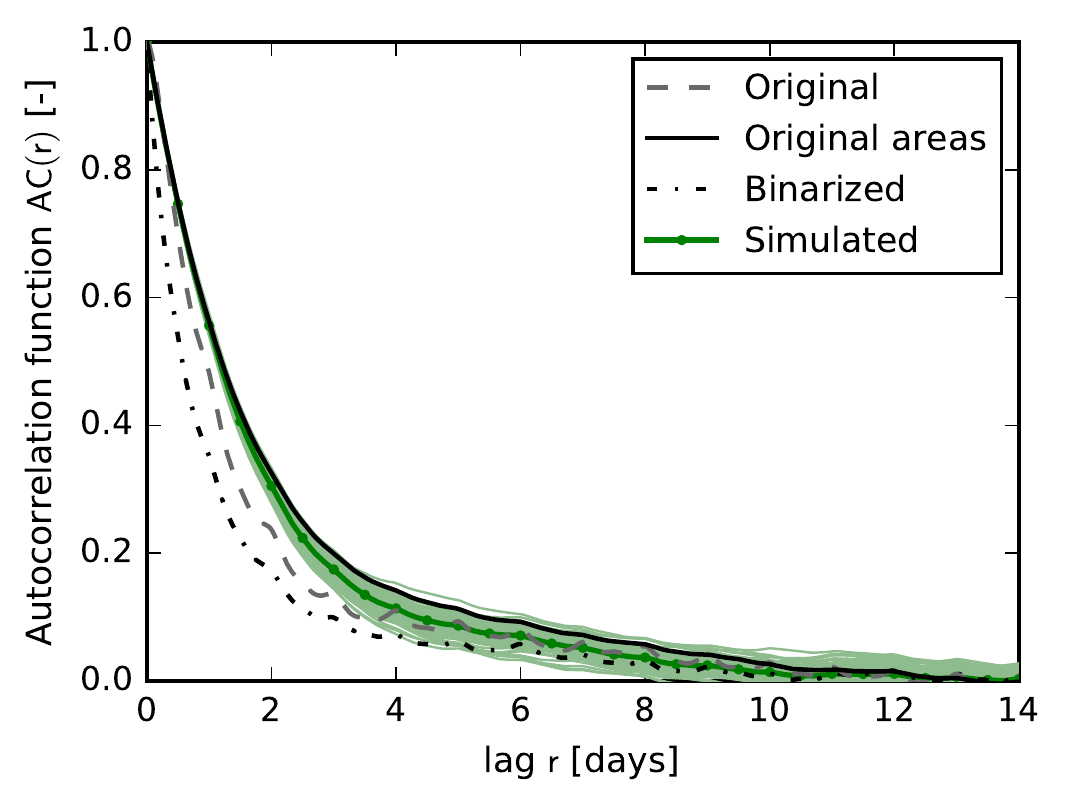}
\caption{
\label{fig:autocorr_delta}
(Color online) Autocorrelation function for the ensemble of 100 simulations using the additive binary Markov chain approach combined with the mismatch energy distribution (green dots) for a memory length of $N = 14$ days. The ensemble average is represented as thick green line. Results have to be compared to those of the piece-wise averaged original renewables.ninja \cite{Staf16} time series (black). Additionally, results for the original time series (grey) and for the binarized time series (black dashes) are shown.  
}
\end{figure}

\subsubsection*{Autocorrelation function}

The autocorrelation of the ensemble of the simulated time series is shown in \cref{fig:autocorr_delta}. As we are dealing with piecewise constant positive and negative values ($A^+(\tau^+)$ and $A^-(\tau^-)$, respectively), we have to compare the autocorrelation functions of the simulations to the autocorrelation function derived from the $A^+$- and $A^-$-values of the original time series (denoted as `Original areas'). We find that the model underestimates the autocorrelation function for all time lags. One reason for this could be that the assignment of mismatch energies is done in a too simplified way. Another reason for the underestimation could be that we tend to underestimate the average mismatch energies, even after the rescaling (cf.~\cref{eq:A_scaling}). Thus, more information on the dependence of $\tau$ and $A$ should be taken into account in order to reconstruct the autocorrelation function more exactly.

\section{Summary and Conclusion} \label{sec:Conclusion}
Temporal correlations of wind power generation are a crucial factor to be considered for the integration of high shares of renewables \cite{Sims11,Bloo16,olau16,Heid10,Rasm12,diaz12,Elsn15,Mila13,Schlachtb16,Bloo16,Cann15,Weber17}. However, it is difficult to model these correlations and simple Markov models often fail to represent them properly \cite{Cara13a,Pesc15,Brok09}.

We introduced a model to accurately reproduce the temporal correlations of wind power generation time series -- focusing on the resting time of binary under- and overproduction events. For this purpose, we used the concept of additive binary Markov chains introduced in \cite{Usat03,Meln06}. This concept allowed us to model binary wind generation data by employing the empirical autocorrelation function of a binarized input time series.

For a wide range of memory lengths, the autocorrelation function, the average backup need as a function of the storage capacity and the resting time distribution of the binarized input time series can be reproduced almost perfectly by the additive binary Markov model. A small memory length better represents short-range correlations whereas a long memory length leads to a better representation of the long-range correlations. Furthermore, we showed, that the exact choice of the renewable penetration $\gamma$ does hardly impact the results.

In order to transform the simulated binary time series back to a non-binary time series, we used the joint probability of the resting time $\tau$ and the mismatch energy $A$. The higher the resting time, the higher the mismatch energy, on average. The application of this relation allowed us to derive the average backup energy as a function of the storage capacity and to compare it directly to that of the original time series. We found that our model is able to reconstruct this function for storage capacities of up to about 100 hours. The autocorrelation function is underestimated, potentially indicating that the relation between $\tau$ and $A$ is not sufficiently well captured.

The presented model is fairly simple to implement and still captures essential features of wind generation time series, especially long memory times. It thus overcomes essential problems of previous approaches \cite{Brok09,Papa08,Hagh13,Pesc15,Cara13a,Cara13}. In the present paper, we demonstrated the ability of the additive binary Markov model for wind time series on national scales. However, it may also be used to capture temporal characteristics in different regimes such as smaller spatial and/or temporal scales. For future applications, this model can be used to learn more about the temporal structure of high- and low-wind periods, for example when planning to combine a wind farm with a storage device.

\section*{Acknowledgments}
We thank J.~Wohland, T.~Pesch, M.~Reyers, T.~Faulwasser, and T.~Brown for stimulating discussions. We gratefully acknowledge support from the Helmholtz Association (via the joint initiative ``Energy System 2050 -- A Contribution of the Research Field Energy'' and the grant no.~VH-NG-1025) and the Federal Ministry for Education and Research (BMBF grant no. 03SF0472) to D.~W..


%

\end{document}